# High Resolution IRAS Maps and IR Emission of M31
## — II. Diffuse Component and Interstellar Dust

**Cong Xu**

Max-Planck-Institut für Kernphysik, Postfach 103980, 69117 Heidelberg, Germany

**George Helou**

IPAC 100-22, California Institute of Technology, Pasadena, CA 91125

July 23, 1995

Send correspondence to C. Xu

# ABSTRACT

Large-scale dust heating and cooling in the diffuse medium of M31 is studied using the high resolution (HiRes) IRAS maps in conjunction with UV, optical (UBV) and the HI maps. A dust heating/cooling model is developed based on a radiative transfer model which assumes a 'Sandwich' configuration of dust and stars and takes fully into account the effect of scattering of dust grains. The model is applied to a complete sample of 'cells' (small areas of size $2' \times 2'$), generated from the above maps. The sample covers the M31 disk in the galactocentric radius range 2 — 14 kpc, and includes only the cells for which the contribution of the discrete sources to the $60\mu m$ surface brightness is negligible ($< 20\%$). This effectively excludes most of the bright arm regions from our analysis. We find that:

(1) The mean optical depth (viewed from the inclination angle of 77°) increases with radius from $\tau_V \sim 0.7$ at $r = 2$ kpc outwards, reaches a peak of $\sim 1.6$ near 10 kpc, and stays quite flat out to 14 kpc, where the signal falls below the $5\sigma$ level.

(2) A correlation between $\tau_V$ and HI surface density is suggested by the similarity between their radial profiles. Significant differences are found between the radial profiles of the $H_2$ gas (estimated from CO) and of the dust (from $\tau_V$), which are most probably due to the large uncertainty in the CO-to-$H_2$ conversion factor, and to the under-representation of $H_2$-rich regions in the sample of cells of *diffuse* regions.

(3) The $\tau_V/N(HI)$ ratio decreases with increasing radius in the disk of M31, with an exponential law fit yielding an e-folding scale length of $9.6 \pm 0.4$ kpc.

(4) The optical depth adjusted for this gradient, $\tau_{V,c}$, is strongly and linearly correlated with N(HI) over one and a half order of magnitude of column density, indicating that at a given radius $r$ the dust column density is proportional to the HI gas column density, with the proportionality factor decreasing with increasing $r$.

(5) With the assumption that the ratio of $\tau_V$ to dust column density is the same as that in Solar Neighborhood, the portion of the M31 disk at radii between 2 and 14 kpc contains $1.9 \pm 0.6\ 10^7\ M_\odot$ of dust, yielding a global dust–to–total-gas mass ratio of $9.0 \pm 2.7\ 10^{-3}$, very close to Solar Neighborhood value.

(6) The non-ionizing UV radiation, mainly due to B stars ($4$–$20 M_\odot$), contributes only 27% of the heating of the diffuse dust in M31. Throughout the M31 disk, heating of the diffuse dust is dominated by optical radiation from stars at least a billion years old.

*Subject headings*: galaxies: individual – galaxies: interstellar matter – galaxies: photometry – interstellar: grains

# 1. Introduction

This is the second paper in a series on the infrared (IR) emission of the Andromeda Galaxy (M31) studied using the new high resolution (HiRes) IRAS maps. In the first paper (Xu & Helou 1995, hereafter Paper I) we studied the overall morphology and the discrete sources of the far-infrared (FIR) emission in the disk of M31. In this paper we investigate the diffuse FIR component and the properties of the interstellar dust in M31.

M31 is an ideal target for studies of the FIR emission of the diffuse interstellar dust *not* associated with star formation regions in a galaxy other than the Milky Way. First of all, being the nearest spiral outside the Milky Way, M31 is well resolved by IRAS. Thus the discrete sources which represent most of the FIR emission associated with star formation regions (Paper I) can be distinguished from diffuse emission. Second, because M31 has a very low present-day star formation rate, about an order of magnitude lower than that of the Milky Way (Walterbos 1988; Paper I), the diffuse component dominates the FIR emission of M31 whereas the dust associated with HII regions contributes only $30 \pm 14\%$ of total IR luminosity of M31 (Paper I).

It has been well established that the diffuse FIR emission in a galaxy is due to the thermal radiation of the interstellar dust heated by the interstellar radiation field (Jura 1982; Cox et al. 1986; Helou 1986; Lonsdale-Persson & Helou 1987; Xu & De Zotti 1989). The diffuse FIR emission has been widely used to study the properties of the interstellar dust, e.g. abundance of dust, dust–to–gas ratio, composition and size distribution of dust grains, heating and cooling of dust grains, etc. (Draine & Anderson 1985; Walterbos & Schwering 1987, hereafter WS87; Désert et al. 1990). Xu & Helou (1994) studied the IRAS color-color diagrams of the diffuse component in the M31 disk and they found rather low $I_{60\mu}/I_{100\mu}$ ratios and high $I_{12\mu}/I_{25\mu}$ ratios for those regions where the interstellar radiation field (ISRF) is low. This was interpreted as evidence of deficiency of Very Small Grains (but not PAH) in the M31 disk. In this paper we attempt to answer the following questions:

(1) How is the interstellar dust distributed in M31? How does it correlate with HI gas and $H_2$ gas?

(2) How much interstellar dust does M31 have? What is the dust–to–gas ratio?

(3) What is the energy budget of the diffuse interstellar dust?

In the literature, the most common approach for studying the abundance and the distribution of dust using its FIR emission is based on the assumption that the dust is in thermal equilibrium with a characteristic temperature $T_d$ (see the review by Soifer et al. 1987). Then the amount of the dust may be estimated from the infrared optical depth defined by $\tau_\lambda = I_\lambda/B_\lambda(T_d)$ (WS87; Deul 1989). There are two problems with this approach: (1) If there are several temperature components sampled along the line of sight, as is likely, the characteristic temperature is biased to the highest values, because of the strong dependence of the emission on temperature (WS87); (2) the assumption that the dust responsible for the FIR emission is in thermal equilibrium may not be valid because, particularly in the low ISRF regions, a substantial part of FIR emission detected by IRAS may be due to grains undergoing temperature fluctuations (Désert et al. 1990). These two biases both tend to overestimate the emissivity and, consequently, underestimate the dust abundance. As a consequence, the dust–to–gas ratio resulting from such an approach is usually on the order of $10^{-3}$ (Devereux and Young, 1990), about an order of magnitude lower than the canonical value estimated from optical studies (Savage and Mathis 1979; van den Bergh 1975).

In this paper we develop a dust heating and cooling model which takes a different approach. The model is based on the assumption that the interstellar dust is heated by the non-ionizing UV (912Å — 3650Å) and the optical (3650Å — 9000Å) radiation in the interstellar radiation field (ISRF), and cools down by infrared radiation ($8 - 1000\mu m$). In other words, the model treats dust grains as 'frequency-converters' which convert the UV and optical radiation to FIR radiation via the absorption-reradiation process. It is from the fraction so converted rather than from the emission alone that we estimate the optical depth of the disk, i.e. the column density of dust. Thus no assumption about the dust thermal equilibrium and $T_d$ is needed. By adopting an empirical extinction curve of M31 (Hutchings et al. 1992; Walterbos and Kennicutt 1988), our model is also insensitive to the grain model (i.e. the composition and the size distribution of dust grains). Quantitatively, our model is based on a radiative transfer code which takes the observed intensities of the FIR, UV, and optical emissions as input. It then estimates (1) the optical depth of the disk which can be converted to dust column density, (2) the amount of heating of dust due to the non-ionizing UV and to the optical radiation respectively, and in principle (3) the extinction-corrected surface brightness of the UV and optical radiation at different wavelengths.

In Section 2 we describe the data. The model for the heating and cooling of interstellar dust is presented in Section 3. The results are given in Section 4. A discussion is carried out in Section 5. Section 6 contains the summary. As in Paper I, we assume for M31 a distance of 690 kpc (1' = 200 pc along the major axis), an inclination angle of $i = 77°$ ($i = 0$ for seen face-on), and P.A. = 37°.

## 2. The data

We define the diffuse FIR component as the emission from interstellar dust not associated with HII regions. Sources at $60\mu m$ (Xu & Helou 1993; Paper I) are used to represent the dust emission associated HII regions. In order to distinguish regions where the sources dominate from those where the diffuse emission dominates, we break the M31 field into small 'cells', each of size $2' \times 2'$, corresponding to a linear size of $0.4 \times 1.8$ kpc$^2$ in the plane of M31. Point-by-point studies are made on these cells. The HiRes maps as well as an UV (2030Å) photometry map (Milliard 1984) which has an angular resolution of about 1'.5, three optical (UBV) photometry maps (Walterbos & Kennicutt 1987) of resolution $\sim 15''$, and a HI map (Brinks 1984; Brinks & Shane 1984) of resolution of $48'' \times 72''$, are smoothed to a common 1'.7 (HPBW) round beam on a grid with 0'.5 pixels. A complete sample of the M31 cells is constructed with these maps, with the surface brightness of each cell taken from the corresponding map by averaging the surface brightness of a 4 × 4 array of adjacent pixels (Xu & Helou 1994). Cells are included only if:

(1) both the 60$\mu m$ and 100$\mu m$ surface brightness exceed a signal-to-noise ratio of 5;

(2) the contribution from the FIR sources is negligible, namely $Q_s = I_{60\mu,s}/I_{60\mu} \leq 0.2$, where $I_{60\mu,s}$ being the contribution of the sources to the 60$\mu m$ surface brightness taken from the source map (Xu & Helou 1993) and $I_{60\mu}$ the total 60$\mu m$ surface brightness;

(3) they are not located in the optical bulge, which is the central ellipse of size $12' \times 20'$ (Walterbos and Kennicutt 1988, hereafter WK88).

The sample contains 358 such cells.

It should be noted that the 60$\mu m$ source map (Xu & Helou 1993) includes all the 70 sources extracted using the Gaussian fitting procedure with the size of fitting area of $4'.5 \times 4'.5$. As pointed out in Paper I, these sources suffer confusion problems, and 16 of them have not been included into the final list of sources in Paper I. On the other hand, the sources under-represent the emission of the HII-region-associated dust (Paper I), hence some cells in the sample may be contaminated by large contribution from this dust. Nevertheless, these uncertainties will have little effect on our conclusions on the diffuse dust since our analysis is based on statistics of a large sample of cells most of which are likely to be dominated by the true diffuse component.

### 3. Model for dust heating and cooling

Applied to each cell ($2' \times 2'$) in the M31 disk, the model takes as input from the observed intensities of the optical, UV and FIR emissions, and gives as output the optical depth to V-band (5500Å) radiation $\tau_V$ by solving the radiative transfer equation. The model also estimates the amount of dust heating due to the UV — optical radiation at different wavelengths. The optical depths to radiation at wavelengths other than 5500Å can be estimated from $\tau_V$ using the adopted extinction curve given in Table 1.

The radiative transfer model is based upon the algorithm developed by van de Hulst & de Jong (1969), which takes the effect of scattering fully into account in the sense that scattered light of any order has been calculated using an iteration procedure from lower order scattered light. An infinite-plane-parallel geometry for the radiative transfer problem is adopted. In contrast with van de Hulst & de Jong (1969), we allow for different thicknesses of the star layer and dust layer. A 'Sandwich' configuration (Fig.1) is used for the dust and star distributions in a galaxy disk with the scale height of the stars assumed to be larger than or equal to that of the dust.

Following van de Hulst & de Jong (1969), the direct star light is defined as of zero-th order:

$$\begin{aligned} I_0(\tau_1, \mu_0) &= S_0(1 - e^{-\tau_1/\mu_0}) + \frac{E_1}{\mu_0} e^{-\tau_1/\mu_0} & (\mu_0 > 0) \\ &= S_0(1 - e^{-(\tau-\tau_1)/(-\mu_0)}) + \frac{E_2}{\mu_0} e^{-(\tau-\tau_1)/(-\mu_0)} & (\mu_0 < 0) \end{aligned} \quad (1)$$

where $I_0$ is the zero order intensity seen at a point (e.g. A in Fig. 1) in the plane toward direction $\mu_0 = \cos(i_0)$; and

$$E_1 = \int_{-z1}^{0} \epsilon \, dz \quad (2)$$

is the integral of the emissivity $\epsilon$ over the layer of stars lying in front of the dust layer, and

$$E_2 = \int_{z0}^{z0+z2} \epsilon \, dz \quad (3)$$

the integral of the emissivity over the layer of stars lying behind the dust layer. The zero order source density $S_0$ is assumed constant within the dust layer. The definitions of $\mu_0$, $\tau_1$, $\tau$, $z_0$, $z_1$, and $z_2$ are illustrated in Fig.1. The first and higher order scattered radiation can be calculated as following

$$\begin{aligned} I_n(\tau_1, \mu_n) &= \int_{\tau_1}^{\tau} S_n(\tau', \mu_n) e^{(\tau_1-\tau')/(-\mu_n)} \frac{d\tau'}{-\mu_n} & (\mu_n < 0) \\ &= \int_{0}^{\tau_1} S_n(\tau', \mu_n) e^{(\tau'-\tau_1)/\mu_n} \frac{d\tau'}{\mu_n} & (\mu_n > 0) \end{aligned} \quad (4)$$

where subscript n ($\geq 1$) denotes the n-th order scattered light. The source density that gives rise to this intensity is

$$S_n(\tau', \mu_n) = \frac{a}{2} \int_{-1}^{1} I_{n-1}(\tau', \mu_{n-1}) \phi(\mu_n, \mu_{n-1}) \, d\mu_{n-1} \qquad (n \geq 1) \qquad (5)$$

where $a$ ($\sim 0.5$) is the albedo and $\phi$ the phase function, both taken from Mathis et al. (1983). The total intensity at any point (e.g. the point A in Fig.1) in the dust disk toward any direction $\mu' = \cos(i')$ is then found from

$$I(\tau_1, \mu') = \sum_{n=0}^{\infty} I_n(\tau_1, \mu') \qquad (6)$$

As pointed out by van de Hulst & de Jong (1969), for small $\tau$ and $a \leq 1$ this series converges rapidly. In practice we took the first 20 terms in the summation. For the adopted albedo and phase function this gives an approximation to Eq(6) with an accuracy of $10^{-5}$ for $\tau \leq 10$.

Using this radiative transfer model, we then calculate the ratio between the light absorbed by dust (dust heating) at a certain frequency $\lambda$ and the light escaped and eventually observed at the same frequency, as a function of the optical depth of the disk at that frequency, $\tau_\lambda$, and the view angle:

$$G(\tau_\lambda, \mu) = \frac{\int_0^{\tau_\lambda} \left[ \int_{4\pi} I_\lambda(\tau_1, \mu') d\omega \right] (1-a) d\tau_1/\mu}{I_\lambda(0, \mu) + E_1/\mu}, \qquad (7)$$

where $I_\lambda(0, \mu) + E_1/\mu$ is the prediction for the light observed, $\mu = \cos(i)$ (for M31 $i = 77°$), $\tau_\lambda$ the optical depth of the disk (seen face on), $\omega$ the solid angle, and $a$ the albedo. The factor $(1-a)$ gives the ratio between the absorption cross section and the extinction cross section. It is worthwhile to note that the resulting value of $G(\tau, \mu)$ is not sensitive to the values of $E_1$ and $E_2$ as long as the disk is optically thin. It should be mentioned that, although not used in this work, the model also calculates the extinction:

$$A_\lambda = 2.5 \times \log \left( \frac{(E_1 + E_2 + S_0 \tau_\lambda)/\mu}{I_\lambda(0, \mu) + E_1/\mu} \right). \qquad (8)$$

We calculate dust heating due to UV light (912Å $< \lambda \leq$ 3000Å) and that due to optical and NIR light (3000 $< \lambda \leq$ 9000Å) separately, and assume that dust heating due to radiation of wavelength $> 9000\text{Å}$ is negligible. The stars most responsible for the non-ionizing UV radiation (B-stars) presumably have similar scale height as that of the dust (Mathis et al. 1983). Therefore for the UV radiative transfer model we assume that both $z_1$ and $z_2$ in Fig.1 are equal to zero, and so the values of $E_1$ and $E_2$ in Eq(1). On the other hand, it is likely that the older stars responsible for the optical radiation have larger scale height than the dust. We assume that stars which lie outside the dust layer radiate half of the total optical radiation, and that $E_1 = E_2 = 0.5 \times S_0 \tau$.

The only UV photometry observations covering the whole M31 field have been made by the Marseille group (Deharveng et al. 1980; Milliard 1984) at 2030Å. Therefore we calculate the UV heating using the formula:

$$I_{fir}^{UV} = I_{2030} \times 10^{0.4 A_{2030}^G} \times G(\tau_{2030}) \int_{912\text{Å}}^{3000\text{Å}} 10^{-0.4(m(\lambda) - m(2030\text{Å}))} \times \frac{G(\tau_\lambda)}{G(\tau_{2030})} \, d\lambda \quad (9)$$

where $I_{fir}^{UV}$ is the FIR radiation intensity due to UV heating; $I_{2030}$ the 2030Å radiation intensity taken from the UV map (Milliard 1984), in units of erg cm$^{-2}$ s$^{-1}$ Å$^{-1}$ arcsec$^{-2}$; and $\tau_{2030}$ the optical depth of the disk at 2030Å. We have omitted the variable $\mu$ in the expression of $G(\tau_\lambda)$ because here $\mu = \cos(77°)$ is fixed. $A_{2030}^G$ is the foreground Galactic extinction at 2030Å. Burstein & Heiles (1984) found that in the direction of M31, the Galactic reddening is $E(B-V)_G = 0.08$ (see also van den Bergh 1991). According to Savage & Mathis (1979),

$$A_{2030}^G = 8.67 \times E(B-V)_G = 0.69 \text{ mag.} \quad (10)$$

The adopted UV spectrum $(m(\lambda) - m(2030\text{Å}))$ and selective UV extinctions $(\tau_\lambda/\tau_V)$ are given in Table 1. Following Koper (1993), the UV spectrum is estimated from the mean of the UV spectra of 29 fields in the M31 disk, each $2'.5 \times 2'.5$ in size, observed by Israel et al. (1986) in five UV bands: 1550, 1800, 2200, 2500 and 3300Å. It is likely that the UV spectrum changes from place to place in a spiral disk, which may introduce an error into our model calculation. However, this is only on the second order compared to the uncertainties due to errors in the absolute flux at 2030Å. And also since the UV heating is never dominant for the diffuse dust in the M31 disk (Section 4.2), this uncertainty will have little effect on our results. The UV extinction curve is taken from Hutchings et al. (1992), who found a much shallower and narrow 2175Å bump compared to the Galactic extinction curve (Savage & Mathis 1979). The magnitude and the selective extinction at 912Å are obtained from extrapolation of the available UV data reported in Table 1.

We have more information about the spectral distribution of the optical radiation field in M31. Digitized photographic surface photometry maps in three optical bands (Walterbos & Kennicutt 1987), namely the U-band ($\lambda_1 = 3650$Å), B-band ($\lambda_2 = 4400$Å), and V-band ($\lambda_3 = 5500$Å), are available to us. The R-band ($\lambda_4 = 7000$Å) magnitude of each pixel in the M31 field can be estimated from the V-magnitude using the average (V−R) color of M31 (WK88):

$$V - R = 0.72 \ , \tag{11}$$

which is rather constant in the disk of M31 without any significant gradient along the galactocentric radius (WK88).

Thus we can calculate the dust heating by the radiation at the four corresponding wavelengths:

$$I_{\rm fir}^{\rm op,i} = I_i \times 10^{0.4 A_i^G} \times G(\tau(\lambda_i)) \tag{12}$$

where $i = 1,2,3,4$ correspond to U,B,V,R bands, respectively. Three of four optical selective extinctions ($\tau(\lambda_i)/\tau_V$, i=1,2,3) are taken from WK88, while selective extinction $\tau_R/\tau_V$ is taken from the local value (Savage & Mathis 1979). They are also listed in Table 1. The other two points, $I_{\rm fir}^{\rm op,0}$ and $I_{\rm fir}^{\rm op,5}$, are extrapolated to $\lambda_0 = 3000$Å and $\lambda_5 = 9000$Å, from the four points described above. The final estimate of the optical heating of the diffuse dust is calculated from:

$$I_{\rm fir}^{\rm op} = \sum_{i=0}^{5} 0.5 \times (I_{\rm fir}^{\rm op,i} + I_{\rm fir}^{\rm op,i+1}) \times (\lambda_{i+1} - \lambda_i) \ . \tag{13}$$

TABLE 1.

UV Spectrum and Selective Extinction of M31

| $\lambda$ | $m_\lambda - m_{2030}$ | $\tau_\lambda/\tau_V$ | ref |
|---|---|---|---|
| (Å) | (mag) | | |
| 1550 | -0.33 | 3.47 | 1 |
| 1800 | -0.15 | 3.06 | 1 |
| 2030 | 0.00 | 2.74 | 1 |
| 2200 | 0.08 | 2.82 | 1 |
| 2500 | 0.28 | 2.14 | 1 |
| 3000 | 0.48 | 1.75 | 1 |
| 3650 | | 1.53 | 2 |
| 4400 | | 1.35 | 2 |
| 5500 | | 1.00 | 2 |
| 7000 | | 0.82 | 3 |

References of selective extinction:

1 Hutchings et al. 1992

2 Walterbos and Kennicutt 1988

3 Savage and Mathis 1979

The sum of the optical heating and UV heating,

$$I_{fir}^{tot} = I_{fir}^{op} + I_{fir}^{UV} \qquad (14)$$

estimates the total heating. In order to compare it with the the FIR surface brightness $I_{fir}$, we have to estimate the fraction of dust reradiation in the wavelength range of (40

— $120\mu m$). Fitting the mean IRAS colors of M31 cirrus (Xu & Helou 1994) with a grain model assuming that the very small grains are only half as abundant in M31 dust as they are in Galactic cirrus (Xu & Helou 1994), we find

$$f = \frac{I(40-120\mu)}{I(8-1000\mu)} = 0.32 \pm 0.06 \ . \tag{15}$$

The one-$\sigma$ error is estimated from the uncertainties of the mean IRAS colors, and does not include the uncertainties of the model. In particular, in cases that the FIR emission is dominated by very cold grains ($T \sim 10K$) which emit almost all of their energy outside IRAS bands, the expression in Eq(15) may over-estimate the fraction. On the other hand, given the very sensitive dependence of dust emission on temperature (Section 1), very cold grains can rarely dominate the dust emission. For grains warmer than 15K ($\nu^2$ emissivity law) expression (15) is accurate within 40%.

In the model of dust heating and cooling, the only free parameter is $\tau_V$, the optical depth at 5500Å. The following equation, which balances the cooling rate (the FIR emission) to the heating rate, determines $\tau_V$:

$$I_{\rm fir} = I_{\rm fir}^{\rm tot}(\tau_V) \times f \ , \tag{16}$$

where $I_{\rm fir}$ is the observed FIR ($40-120\mu m$) surface brightness.

## 4. Results

### 4.1. Optical depth and optical-depth–to–gas ratio

A gray-scale plot of the distribution of the V-band optical depth seen from 77° inclination angle, $\tau_V$, calculated using the above model for the cells selected, is presented in Fig.2. The range of the gray-scale is between $\log(\tau_V) = -0.4$ ($\tau_V = 0.4$) and $\log(\tau_V) = 0.6$ ($\tau_V = 4$). Because of the selection criteria of the cells (see Section 2), not the entire disk of M31 is covered in this plot, in particular the star formation regions as indicated by discrete FIR sources have been deliberately left out. There are several spots where the dark cells ($\tau_V \gtrsim 2$) are concentrated, most of them in the 'ring'. There is a general trend that in the inner disk the optical depth tends to be relatively low, and in the ring the optical depth tends to be high.

In Fig.3 we plot $\tau_V$ versus the galactocentric radius for the same sample of M31 cells. The large solid squares with error bars are the means and dispersions for cells in different radius bins, each spanning 2 kpc. The optical depth increases with radius from $\tau_V \sim 0.7$ at r = 2 kpc in the inner part of the disk, reaching a peak of $\tau_V \sim 1.6$ at about 10 kpc, and staying quite flat out to 14 kpc. Beyond 14 kpc no cells are detected in FIR above the $5\sigma$ threshold. This $5\sigma$ threshold might bias the mean values of the optical depth toward higher values, in particular for the inner-most and the outer-most bins where the low FIR emission regions are most likely be found, because the low FIR surface brightness cells generally have low $\tau_V$. In order to check this possibility, we have calculated the means of $\tau_V$ for the same bins using another sample including all cells with signal–to–noise ratio of the FIR surface brightness above a factor of 2. No significant difference is found compared to the means plotted in Fig.3, although the dispersions are increased because of the larger errors in the data. Hence, the bias mentioned above is not significant. For cells in the immediate vicinity of massive star formation regions there is another source of uncertainty, namely that the dust heating model (Section 3) ignores heating by ionizing radiation. This results in an over-estimate of the optical depth for these cells. However, because the sample were selected against such cells which in any case occupy only a small part of the M31 disk, this error shall not affect the means significantly.

It should be noted that the overall distribution of $\tau_V$ is rather flat, which disagrees with a popular assumption (Disney et al. 1989; Byun et al. 1994) that the optical depth and stars have the same central-peaked exponential r-distribution in spiral disks. Another point we want to make is that the face-on optical depth in the disk of M31 should be a factor of 4.4 lower than $\tau_V$ which is seen from 77° inclination angle. The results presented in Fig.3 then indicate that in the M31 disk the face-on optical depth is in the range of 0.2 ∼ 0.4, namely that the disk is optically thin, inconsistent with the hypothesis that spiral disks are "opaque" (Valentijn 1990).

In Fig.4 we compare our results, for which the error bars now represent the statistical uncertainty of the means (uncertainty = dispersion/$\sqrt{N-1}$), with the distributions of the neutral atomic gas (Brinks & Shane 1984) and of the molecular hydrogen gas estimated by Koper et al. (1991) from the CO observations, and that of the dust clouds (Hodge 1980). It appears that the $\tau_V$ distribution shows a similar $r$-dependence as the HI gas distribution, while the molecular gas distribution is very different, characterized by two very narrow peaks at $r \sim$ 4kpc and at ∼10kpc, respectively. This result suggests a correlation between column densities of the HI gas and the diffuse dust. However, the $\tau_V$ vs. $H_2$ comparison must be taken with caution. First of all, regions where the molecular gas is concentrated are under-represented in the sample of cells due to the exclusion of cells dominated by discrete sources (active star formation regions). And second, there seems to be large uncertainties with the $H_2$ map deduced from the CO survey of Koper et al. (1991), in particular in the inner part of the disk. Allen & Lequeux (1993; see also Allen et al. 1994) detected very cold (close to the 2.7 K cosmic background temperature) CO emission in two dust clouds in the inner region of M31 ($r \lesssim$ 2–3 kpc) which may contain as much as a few times $10^7 M_\odot$ molecular gas, and may have a CO to $H_2$ conversion factor of about one order of magnitude higher than the average value taken by Koper et al. (1991). The surface density of dust clouds decreases rapidly and monotonically with increasing radius, perhaps due to the selection effect that clouds close to the center of the galaxy are easily detected on an optical image (Hodge 1980) because of the high brightness of the disk.

We also plotted the results of Hodge & Lee (1988), denoted by the open squares with error bars, from a reddening study for stars (most in OB associations) in six M31 fields with different galactocentric radii. Assuming that the stars are located in the middle of the dust layer, the reddening (E(B − V)) reported in that paper has been converted to $\tau_V$ by multiplying a factor of 2 × 0.921 × 2.8 (WK88). In spite of the fact that completely different approaches are used, a good general agreement is found between the results of Hodge & Lee (1988) and ours, with the former being only slightly higher where the two data sets overlap. The difference may be due to their fields being biased for star formation regions, while ours biased against such regions.

In Fig.5 we plot for the sample of M31 cells the optical-depth–to–HI-gas ratio versus the galactocentric radius. The dashed line represents the optical-depth–to–HI-gas ratio in Solar Neighborhood (Savage & Mathis 1979). It appears that the ratio in M31 is in general slightly higher than that in Solar Neighborhood. A clear trend of decrease of the ratio with increasing radius is evident. The linear regression of the plot gives

$$\log\left(\frac{\tau_V}{N(HI)}\right) = 0.41(\pm 0.02) - 0.045(\pm 0.002) \times \left(\frac{r}{1\,\text{kpc}}\right) \quad . \tag{17}$$

where N(HI) is in units of $10^{21}$ atoms/cm$^2$. The slope ($-0.045\pm 0.002$) corresponds to an $e$-folding scale length of $9.6\pm 0.4$ kpc for the galactocentric gradient of the optical-depth–to–HI-gas ratio in M31. It is interesting to note that our optical-depth–to–HI-gas ratio gradient is consistent with the metallicity gradient found by Blair et al. (1982), which ranges from 9.8kpc (for the N/H abundance ratio) to 14.9kpc (for the O/H abundance ratio). The solid squares with error bars in Fig.5 are the mean ratios of cells in different radius bins, each 2 kpc wide. The error bars represent 1-$\sigma$ dispersions, and amount only to about 30%. The dotted curve is a smooth fit to the means which can be expressed as

$$\log\left(\frac{\tau_V}{N(HI)}\right) = 0.58 - 0.08 \times \left(\frac{r}{1\,\text{kpc}}\right) + 0.22 \cdot \exp\left(\frac{r - 12\,\text{kpc}}{3\,\text{kpc}}\right) \quad (r \leq 14\text{kpc}). \tag{18}$$

In Fig.6, we compare our results on optical-depth–to–HI-gas ratio with those from optical studies of M31. Bajaja & Gergeley (1977) compared the reddening of 121 M31 globular clusters with the HI column density obtained by the Cambridge 21cm line survey (Emerson 1974). Their results, after converting E(B-V)/HI to $\tau_V$/HI assuming $\tau_V/E(B-V) = 0.921 \times 2.8$ (WK88) and $p = 0.5$ (half of the globular clusters are behind the plane of the galaxy), are plotted in Fig.6 (crosses with error bars), and appears to be in reasonably good agreement with ours. WK88 estimate the optical extinction in the M31 disk by comparing the optical surface brightness of the near-side half to the far-side half of the disk. Their results, also plotted in Fig.6, fall below our estimates and those of Bajaja & Gergeley (1977) possibly because, as suggested by WK88, they are only lower limits of true values. It should be noted that the real difference between our results and those of WK88 could be even larger, because in their calculation of $\tau_V/HI$ they have used the 'disk HI' map of M31 (Brinks & Burton 1984), derived from the total HI map (Brinks & Shane 1984) under the assumption that the HI disk of M31 is warped (Brinks & Burton 1984). The 'warp model' has been questioned by Braun (1991) who proposed alternatively a 'sticking-out-arm-segments' model to interpret the different HI velocity components. We have used in this study the map of the total HI column density without the disk/warp decomposition.

We introduce a gradient-corrected optical depth:

$$\tau_{V,c} = \tau_V \times \exp\left(\frac{r}{9.6\text{kpc}} - 1\right) \tag{19}$$

which we plot against HI column density in Fig.7, with different symbols denoting cells in the inner-disk (r < 7 kpc), in the 'ring' (7 ≤ r ≤ 11 kpc), and in the out-disk (r < 7 kpc). A strong and linear correlation is found in this log-log plot: the correlation coefficient is 0.89. The dispersion of the logarithm of $\tau_{V,c}/N(HI)$ ratio is 0.12, i.e. 32% on a linear scale. The slope of the least-square-fit shown by the solid line is 1.01±0.02. These results suggest that for a given galactocentric radius the dust column density, as indicated by $\tau_V$, scales closely and linearly with the column density of HI gas, while the scaling factor decreases with increasing galactocentric radius, reflecting the gradient in the dust–to–HI-gas ratio. Tight local correlation between dust and HI gas has been detected in Solar Neighborhood (Savage & Mathis 1979; Boulanger & Pérault 1988).

It should be noted that taking into account the molecular gas component may affect the estimated gradient of the optical-depth–to–gas ratio expressed in Eq(17). In particular the high mean ratios in the first two bins in Fig.5 might indeed be due to the large amount of very cold molecular gas hiding in the inner part of the M31 disk, suggested by the observations of Allen & Lequeux (1993). And some of the scatter in Fig.5 may also be due to the neglect of the molecular gas. On the other hand, we would recall that (1) globally speaking it is likely that the HI phase contains the bulk of interstellar gas in the M31 disk (Koper et al. 1991); (2) a large percentage of the dust associated with molecular gas should concentrate in star formation regions (Cox & Mezger 1989), and its contribution could therefore have been greatly reduced, though not completely removed, by our removal of discrete sources.

4.2. Total dust mass and global dust–to–gas ratio

In order to convert $\tau_V$ to dust column density, we assume that in the M31 disk the ratio of $\tau_V$ to dust column density is the same as that in Solar Neighborhood (Désert et al. 1990), namely that dust in M31 has the same opacity as dust in Solar Nieghborhood. This is plausible because in the M31 disk the composition and size distribution of the large normal grains which dominate the dust mass (Draine & Lee 1984) may be similar to those in Solar Neighborhood, given that the optical extinction is mainly due to large grains and that the optical extinction curve of M31 is similar to that in Solar Neighborhood (WK88). Thus using the HI map (Brinks & Shane 1984) and the mean $\tau_V$–to–HI-gas ratios for different radial bins, we estimate that there is $1.9 \; 10^7 \; M_\odot$ dust, with an error in the order of 30% including the uncertainties due to the data and to the model calculation, within the M31 disk between r = 2kpc and r = 14kpc.

Within the same radius range we estimate that, using the HI data Brinks & Shane (1984), there is $1.8 \; 10^9 \; M_\odot$ HI gas and, according to Koper et al. (1991), $2.5 \; 10^8 \; M_\odot$ 'warm' $H_2$ gas. The amount of very cold $H_2$ such as found by Allen & Lequeux (1993) in the inner disk is not known. If this component is confined to galactocentric radii $< 3 \sim 4$kpc with a surface density $\sim 3 \; M_\odot$ pc$^{-2}$ (Allen & Lequeux 1993), a rough estimate yields a mass of $\sim 10^8 \; M_\odot$ . Therefore in the M31 disk the global dust–to–gas ratio is $9.0(\pm 2.7) \; 10^{-3}$, indeed very close to Solar Neighborhood value of $7.3 \; 10^{-3}$ (Désert et al. 1990).

### 4.3. Energy budget of the diffuse FIR emission

In our model it is assumed that the diffuse interstellar dust is heated by ISRF in non-ionizing UV (912 — 3000Å) and optical-NIR (3000 — 9000Å) bands. In principle different stellar populations can be held responsible for the ISRF in these bands: the UV radiation is mainly due to B stars (4–20 $M_\odot$) which live only $\lesssim 10^8$ years, and optical-NIR radiation due to less massive, older stars ($\gtrsim 10^9$ years). In this sense, the question of which stellar population is most responsible for the heating of the diffuse dust in M31 can be investigated by estimating the relative contributions of the radiation in these different bands to the dust heating.

We apply the dust heating model (Section 3) to a set of galactocentric annuli, each of 6' width (1.2 kpc) in the M31 plan, covering from 2 kpc to 14 kpc. The average optical depth of each annulus is estimated from the average HI column density, assuming the radius dependence of the optical-depth–to–HI-gas ratio as specified by Eq(18). In Fig.8 the solid line is the radial distribution of the FIR surface brightness, and the dashed line the diffuse FIR emission predicted by the heating model. The significant difference between them around the ring and in the outer region of the disk is due to the discrete sources, most of which are in these parts of the disk (Fig.3 in Paper I). Otherwise the difference is less than 10%, well within the uncertainty of the heating model. This is expected because the optical-depth–to–HI-gas ratio adopted here is estimated from the FIR vs. heating comparison for the diffuse dust (Section 4.1). The non-ionizing UV radiation contributes only 27% of the total heating, represented by the dotted-dashed curve in Fig.8. It is most prominent at the ring, but never dominant. Throughout the M31 disk, the optical radiation, shown as the dotted curve, dominates the heating of the diffuse dust. This indicates that the diffuse FIR emission of M31 is mainly due to heating by the optical radiation from relatively old stars, i.e. stars older than a few $10^9$ years. Xu (1990) found that on average the non-ionizing UV radiation contributes 56 — 76% of heating of diffuse dust in spiral galaxies. The UV heating of diffuse dust in the M31 disk appears to be much less significant than for a more typical spiral galaxy (e.g. the Milky Way). This is due to the very low 2000Å–to–blue flux ratio of M31, actually one of the lowest in a large sample of nearby galaxies (Buat & Xu 1995), a consequence of the very low recent star formation rate of M31.

## 5. Discussion

5.1. Uncertainties introduced by assumptions in the model

As discussed in Section 3, our model for dust heating and cooling is insensitive to the assumptions on the physical conditions of dust (e.g. thermal equilibrium and $T_d$) and to the grain model which is still poorly known (Désert et al. 1990). The only parameter affected by these factors is $f$ in Eq.(16), which estimates the fraction of dust emission in the wavelength range $40 - 120\mu m$. Since our result on optical depth depends on $f$ at most linearly, whereas in optically thick cases only logarithmly, we argue that the uncertainty introduced through $f$ cannot be very large. On the other hand, there are some other uncertainties due to various assumptions made in our model, all for the sake of simplicity, which we discuss in this section.

We have implicitly assumed in the model that each cell ($0.4 \times 1.8$ kpc$^2$) is independent of the others, which is strictly not true because stars in neighboring cells can contribute significantly to the heating of dust in a given cell since the photon mean free-path is usually larger than the size of the cells. This is a reasonable assumption however as long as the physical conditions do not change abruptly between adjacent cells, and it is likely to hold in most cases at least for the optical emission which dominates the dust-heating (Section 4.3), given the rather smooth distribution of the optical surface brightness (Walterbos and Kennicutt 1987).

How reliable is the homogeneity assumption for the dust distribution? Hodge (1980) catalogued dust clouds of different sizes throughout the M31 disk, and Allen & Lequeux (1993; see also Allen et al. 1994) reported very cold and dense molecular clouds associated with two of these dust clouds (D268 and D487) in the inner disk. If the interstellar dust is concentrated in these dust clouds, and if the clouds are highly optically thick, then the radiative transfer problem in the M31 disk can be very different from our model of a uniformly distributed dust plane. Hodge & Kennicutt (1982) found that M31 dust clouds are in general optically thin. This is confirmed by recent study of Sofue & Yoshida (1993) who studied the reddening of a complex of M31 dark clouds, including D382, D384 and D395 in Hodge's catalogue. The peak reddening is only $E(B-V) \simeq 0.2$, corresponding $A_V \sim 0.6$ mag. Allen et al. (1994) estimated that the clouds they studied for very cold molecular gas causes a moderate extinction ($A_V = 2\pm 1$ mag). As demonstrated in the studies of Boissé (1990) and Hobson & Scheuer (1993) (see also Caplan & Deharveng 1986), clumpy media with the same amount of dust will always cause less extinction with a flatter extinction curve than homogeneous media do. The optical extinction law in M31 found by WK88 (see also Hodge and Kennicutt 1982) is not very different from that of the diffuse dust in Solar Neighborhood (but also see Iye and Richter 1985). We therefore conclude that most of the diffuse dust not associated with star forming regions in the M31 disk is optically thin and our analysis is affected only to second order by inhomogeneities. However it should be noted, as pointed out by Allen et al. (1994), that the dust clouds themselves may be very clumpy, containing very high density regions ($A_V \sim 10$ mag) with very small filling factors. Dust associated with these tiny dark clumps will be largely missed by our model. While the possibility of missing the coldest dust adds to the uncertainty of the $\tau_V$ profile we derive, it has less of an impact on the estimated dust–to–gas ratio, precisely because the estimation of $\tau_V$ attaches greater weight to the HI component of the interstellar medium.

Another uncertainty may be due to the variation of the relative location of the dust layer and the stellar disk. In our model we adopted a 'Sandwich' configuration where the dust layer is in the middle of the stellar disk. In a dynamical analysis, Braun (1991) found that several arm segments of HI gas stick out of the disk as far as $\sim 1$ kpc. If the dust is mixed with HI in these arm segments, they should be foreground or background relative to the stars. If these arms are optically thin, the uncertainty caused by this on our dust heating model is not very significant, because the G ratio in Eq(7) is then approximately proportional to the optical depth regardless of the location of the dust layer. On the other hand, if they are very optically thick, this can cause dramatic effect because the UV and optical radiation can be totally extinguished when the layer is foreground, or extinction free when the dust layer is background. The disk of M31, viewed at 77° inclination angle, is marginally optically thick ($\tau_V \sim 1$, see Section 4.1). We therefore checked for evidence of such geometric effect.

In Fig.9 we plot the optical-depth–to–HI ratios of M31 cells in the north-west half (near-half) and in the south-east half (far-half) with different symbols as described in the legend. The large solid and open squares with error bars (statistical uncertainties) are means of cells in different radius bins (2 kpc in width), for the north-west half (near-half) and the south-east half (far-half) respectively. There is a systematic difference between the dust–to–HI ratios of the two halves in the sense that the cells in the near-half tend to have higher ratios and those in the far-half lower ratios. This difference is statistically significant (at $\sim 3\sigma$ level) in the three bins $6 \leq r < 8$ kpc, $8 \leq r < 10$ kpc, and $10 \leq r < 12$ kpc, where the 'ring' is encompassed, although the absolute values of the difference are never very large (at $\sim 20\%$ level). This difference might reflect real changes of the dust properties in the two halves. However it may also be artificial if the 'Sandwich' configuration assumed in our model is violated. Running a model which is otherwise the same as that in Section 3 except that the position of the dust layer is adjustable, we found that the near-half/far-half difference can be accounted for if, in the galactocentric radius range encompassing the 'ring' (7 kpc < r < 12 kpc), the dust layer in the near-half is displaced with respect to the stellar disk toward us by a distance of half its thickness, or the dust layer in the far-half is displaced away from us by about the same distance, or the dust layers in both halves are displaced by a distance of one third of their thickness toward and away from us respectively. Interestingly, these possible deviations of the dust layer from the middle plane in the two halves of M31 are consistent qualitatively with the HI warp reported by Braun (1991). No significant differences on the galactocentric gradient of the optical-depth–to–HI-gas ratio and on the total dust content of M31 are introduced by these possible displacements.

## 5.2. FIR vs. UV-optical comparison as a tool for studying extinction and dust-to-gas ratio in galaxies

The dust heating and cooling model presented in this paper can be used to address the widely debated problem about the extinction in disk galaxies (Disney et al. 1989; Valentijn 1990). Except for a few Local Group galaxies, including the Magellanic Clouds (Koornneef 1982; Fizpatrick 1986), M31 (Hodge & Lee 1988; Bajaja & Gergeley 1977) and M33 (Humphreys et al. 1990; Bianchi et al. 1991) where individual stars or star clusters can be resolved, direct reddening/extinction studies through optical observations are not possible for galaxies in general. Indirect methods such as the statistical analysis of the inclination dependence of the surface brightness and the isophotal diameter lead to controversial results (Holmberg 1958; Valentijn 1990; Burstein et al. 1991), possibly due to various selection effects (Disney 1992). Block et al. (1994) proposed that optical minus NIR colors (B−K and V−K) can be used as extinction indicators. But the quantitative result of the method depends sensitively on the assumed intrinsic colors which change with the stellar population. In principle the extinction can be estimated from the dust column density for which the FIR and sub-mm surface brightness might be used as an indicator. However, because of the sensitive dependence of the dust emission model on the temperature and on the poorly known FIR emissivity law of grains, the column density of dust, and therefore the optical depth, estimated from the emission of dust grains alone is very uncertain (Kwan & Xie 1992; Chini & Krügel 1993).

In our model, the FIR/UV-optical ratio has been used as an extinction indicator, under the reasonable assumption that all the radiation absorbed by dust in UV and optical will be re-radiated in the infrared. Quantitative results are obtained using a radiative transfer model (Section 3). Good agreements with optical studies of extinction in M31 (Hodge & Lee 1988; Bajaja & Gergeley 1977; WK88) are found. Such a model can be easily extended to other disk galaxies for which the FIR, UV and optical data are available. Indeed Xu & Buat (1995) made such a study for a sample of 135 nearby spiral galaxies and found that most of spirals in their sample are optically thin to blue radiation ($\tau_B < 1$).

The results shown in Section 4 also demonstrate that our model can be used to analyze quantitatively the dust–to–gas ratio in galaxies. The dust–to–gas ratios estimated using FIR and sub-mm emission of dust (Devereux & Young 1990; Rowan-Robinson 1992; Franceschini & Andreani 1995) are usually significantly lower than that found in Solar Neighborhood, probably due to missing cold ($\lesssim 15K$) dust (Block et al. 1994). Using our model one estimates the dust column density from the extinction rather than from the emission. Since the extinction, unlike the emissivity, does not depend on grain temperature, our method in principle (particularly when sub-mm data are also available) can detect all dust grains no matter how cold they are, unless they hide in clumps of very high optical depth (Boissé 1990). This is illustrated in Fig.10, in which we plot for our sample of M31 cells the diagram of dust–to–HI-gas ratio estimated using our model versus the dust–to–HI-gas ratio estimated from the FIR emission by assuming that (1) the grains are in thermal equilibrium and (2) the temperature of grains are specified by the $60\mu m$ to $100\mu m$ flux ratio ($\nu^2$ emissivity law). Both ratios are normalized by the value in Solar Neighborhood. Although there is a good correlation in the plot, the dust–to–gas ratio estimated from the emission ($M_{d,100}/M(HI)$) is always significantly lower than that estimated from our model ($M_{d,V}/M(HI)$). In particular in the outer part of the disk where the intensity of the ISRF is low, and consequently the FIR surface brightness is also low, the infrared-radiating dust accounts only a few percent of the dust estimated from our dust heating/cooling model.

## 6. Summary

We have investigated the large-scale dust heating and cooling in the diffuse medium of M31 using far-infrared maps from IRAS in conjunction with the UV 2000Å map by Milliard (1984), UBV maps by Walterbos & Kennicutt (1987), and the HI map by Brinks (1984). We analyze the areas of the M31 disk where discrete sources contribute $\leq 20\%$ of the local $60\mu m$ surface brightness, combining the cooling brightness (far-infrared) with the emerging portion of the heating brightness ($912\text{Å} \leq \lambda \leq 9000\text{Å}$) to derive a mean optical depth $\tau_V$ for each cell ($\sim 0.4 \times 1.8 \,\text{kpc}^2$) in the M31 disk. This derivation is based on a plane-parallel radiative transfer model allowing for different thicknesses of the star layer and dust layer ('Sandwich' model), which estimates heating input in each of 12 wavelength intervals, and accounts adequately for both absorption and scattering of light by dust. We find that:

1). The mean optical depth (viewed from the inclination angle of 77°) increases with radius from $\tau_V \sim 0.7$ at $r = 2$ kpc outwards, reaches a peak of $\tau_V \sim 1.6$ near 10 kpc, and stays quite flat out to 14 kpc, where the signal falls below the $5\sigma$ level. Our results are consistent with earlier studies on stellar reddening, and improve on them in spatial coverage and accuracy.

2). Compared to the radial profile of the HI gas and that of the molecular gas, the optical-depth distribution resembles the former but significantly differs from the latter. This suggests a correlation between the HI gas and dust. However, the dissimilarity between the distributions of $H_2$ gas and of optical-depth is probably due to the large uncertainty of the CO-to-$H_2$ conversion factor, and to the fact that $H_2$-rich regions are under-represented in the sample of cells of *diffuse* regions.

3). The $\tau_V/N(HI)$ ratio decreases with increasing radius in the disk of M31, with an exponential law fit yielding an e-folding scale length of $9.6 \pm 0.4$ kpc. On average the $\tau_V/N(HI)$ ratio in M31 is not very different from its value in Solar Neighborhood.

4). The optical depth adjusted for that radial gradient, i.e. $\tau_{V,c} = \tau_V \times (\exp(\frac{r}{9.6\,\text{kpc}}) - 1)$ is strongly and linearly correlated with N(HI) over one and a half order of magnitude of column density. This indicates that at a given radius r the dust column density is proportional to the HI gas column density, with the proportionality factor decreasing with increasing r.

5). The portion of the M31 disk at radii between 2 and 14 kpc contains $1.8\ 10^9\ M_\odot$ of HI gas and $2.5\ 10^8\ M_\odot$ of $H_2$. Using the means of $\tau_V/N(HI)$ ratio at different galactocentric radii in the M31 disk and the dust opacity in Solar Neighborhood, we derive from the HI map a corresponding total dust mass of $1.9 \pm 0.6\ 10^7\ M_\odot$, yielding a global dust–to–total-gas mass ratio of $9.0 \pm 2.7\ 10^{-3}$. This value is about an order of magnitude larger than estimates based on emissivities and temperatures derived from 60-to-100$\mu m$ color ratios. We consider these smaller estimates less reliable because they are severely affected by the inadequacy of the single temperature assumption.

6). The non-ionizing UV radiation, mainly due to B stars ($4-20 M_\odot$) contributes only 27% of the heating of the diffuse dust in M31. This contribution is never locally dominant, but is most prominent at the ring defined by HII regions and molecular clouds maxima. Throughout the M31 disk, heating of the diffuse dust is dominated by optical radiation from stars at least a billion years old.

*Acknowledgements.* We are very grateful to Dr. E. Brinks for providing the HI map of M31, to Dr B. Milliard for the UV (2000Å) map, and to Dr. R. Walterbos for the optical photometry maps. We are indebted to the referee, Dr. R. Walterbos, whose comments helped to improve this paper in various aspects. Helpful discussions with C. Beichman, R. Beck, E. Berkhuijsen, J. Fowler, J. Kirk, and J. Lequeux, and useful exchanges with U. Lisenfeld are acknowledged. Part of the work was done when CX was at Max-Planck-Institut für Radioastronomie, supported by an Alexander von Humboldt Fellowship. He thanks Prof. R. Wielebinski for his hospitality. This research is supported in part through the IRAS Extended Mission Program by the Jet Propulsion Laboratory,

# References

Allen, R.J., Lequeux, J., 1993, ApJ 410, L15

Allen, R.J., Le Bourlot, J., Lequeux, J., Pineau des Forets, G., Roueff, E., 1994, preprint

Bajaja, E., Gergeley, T.E., 1977, A&A 61, 229

Bianchi, L., Hutchings, J.B., Massey, P., 1991, A&A 249, 14

Blair, W.P., Kirshner, R.P., Chevalier, R.A., 1982, ApJ 254, 50

Boissé, P., 1990, A&A 228, 483

Boulanger, F., Pérault, M., 1988, ApJ 330, 964

Block, D.L., Witt, A.N., Grosbol, P., Stockton, A., Moneti, A., 1994, A&A 288, 383

Braun, R., 1991, ApJ 372, 54

Brinks, E., 1984, Ph.D. Thesis, University of Leiden

Brinks, E., Burton, W.B., 1984, A&A 141, 195

Brinks, E., Shane, W.W., 1984, A&AS 55, 179

Buat, V., Xu, C., 1995, A&A, in press

Burstein, D., Haynes, M.P., Faber, S.M., 1991, Nature 353, 515

Burstein, D., Heiles, C., 1984, ApJS 54, 33

Byun, Y.I., Freeman, K.C., Kylafis, N.D., 1994, ApJ 432, 114

Caplan, J., Deharveng, L., 1986, A&A 155, 297

Chini, R., Krügel, E., 1993, A&A 166, L8

Cox, P., Krügel, E., Mezger, P.G., 1986, A&A 155, 380

Cox, P., Mezger, P.G., 1989, A&AR 1, 49

Deharveng, J.M., Jakobsen, P., Milliard, B., Laget, M., 1980, A&A 88, 52

Désert, F.A., Boulanger, F., Puget, J.L., 1990, A&A 273, 215

Devereux, N.A., Young, J.S., 1990, ApJ 359, 42

Deul, E.R., 1989, A&A 218, 78

Disney, M., 1992, Nature 356, 114

Disney, M., Davies, J., Phillipps, S., 1989, MNRAS 239, 939

Draine, B.T., Anderson, N., 1985, ApJ 292, 494

Draine, B.T., Lee, H.M., 1984, ApJ 285, 89

Emerson, D.T., 1974, MNRAS 169, 607

Franceschini, A., Andreani, P., 1995, ApJ, in press

Helou, G., 1986, ApJ 311, L33

Helou, G., Khan, I.R., Malek, L., Beohmer, L., 1988, ApJS 68, 151

Hobson, M.P., Scheuer, P.A., 1993, MNRAS 264, 145

Hodge, P., 1980, AJ 85, 376

Hodge, P., Kennicutt, R., 1982, AJ 87, 264

Hodge, P., Lee, M., 1988, ApJ 329, 651

Holmberg, E., 1958, Medn. Lunds Astr. Obs. 2, 136

Humphreys, M., Bowyer, S., Martin, C., 1990, AJ 99, 84

Hutchings, J.B. et al., 1992, ApJ 400, L35

Israel, F.P., De Boer, K.S., Bosma, A., 1986, A&AS 66, 117

Iye, M., Richter, O., 1985, A&A 144, 471

Jura, M., 1982, ApJ 354, 70

Koornneef, J., 1982, A&A 107, 247

Koper, E., 1993, Ph.D. Thesis, University of Leiden

Koper, E., Dame, T.M., Israel, F.P., Thaddeus, P., 1991, ApJ 383, L11

Kwan, J., Xie, S., 1992; ApJ 398, 105

Lonsdale-Persson, C.J., Helou, G., 1987, ApJ 314, 513

Mathis, J.S., Mezger, P.G., Panagia, N., 1983, A&A 128, 212

Milliard, B., 1984, These, d'Etat, University de Marseille

Rowan-Robinson, M., 1992, MNRAS 258, 787

Savage, B.D., Mathis, J.S., 1979, ARA&A 17, 73

Sofue, Y., Yoshida, S., 1993, ApJ 417, L63

Soifer, B.T., Houck, J.R., Neugebauer, G., 1987, ARA&A 25, 187

Valentijn, E.A., 1990, Nature 346, 153

van de Hulst, H.C., de Jong, T., 1969, Physica 41, 151

van den Bergh, S., 1975, A&A 41, 53

van den Bergh, S., 1991, PASP 103, 1053

Walterbos, R.A.M., 1988, in *Galactic and Extragalactic Star Formation*, eds. R.E. Pudritz and M. Fich, Kluwer, Dordrecht, p.361

Walterbos, R.A.M., Kennicutt, R.C., 1987, A&AS 69, 311

Walterbos, R.A.M., Kennicutt, R.C., 1988, A&A 198, 61 (WK88)

Walterbos, R.A.M., Schwering, P.B.W., 1987, A&A 180, 27 (WS87)

Xu, C., 1990, ApJ 365, L47

Xu, C., Buat, V., 1995, A&A 293, L65

Xu, C., De Zotti, G., 1989, A&A 225, 12

Xu, C., Helou, G., 1993, in *Science with High Spatial Resolution Far-Infrared Data*; eds. S. Terebey, J. Mazzarella, (Pasadena: Jet Propulsion Laboratory), p.87

Xu, C., Helou, G., 1994, ApJ 426, 109

Xu, C., Helou, G., 1995, ApJ, in press (Paper I)

FIGURE CAPTIONS

**Figure 1.** Schematic illustration of definitions of several parameters used in the dust heating model. $\epsilon$ denotes the volume emissivity, $\kappa$ the volume absorptivity, $\mu_0 = \cos(i_0)$, and $\tau_1 = \int_0^{l_1} \kappa \, dz$.

**Figure 2.** Gray-scale plot of $\tau_V$, the V-band (5500Å) optical depth viewed from the inclination angle $i = 77°$, in the M31 disk. Areas containing considerable contribution from discrete sources ($I_{60\mu,s}/I_{60\mu} > 0.2$) are not included. The range of the gray-scale is between $\log(\tau_V) = -0.4$ (the faintest) and $\log(\tau_V) = 0.6$ (the darkest), and the intermediate gray levels correspond linearly to the $\log(\tau_V)$ levels.

**Figure 3.** Radial distribution of $\tau_V$. Crosses are results for M31 cells (size of $2' \times 2'$). Solid squares with error bars are means for cells in six bins, each spanning 2 kpc in the plane of M31. The error bars give the one-$\sigma$ dispersions.

**Figure 4.** Comparison of the radial distribution of optical depth $\tau_V$ with the radial distributions of the HI (Brinks 1984), of the $H_2$ gas estimated from CO (Koper et al. 1991), and of dust clouds (Hodge 1980). The column densities of gas and the column density of dust clouds are in arbitrary units. Plotted are also the results of star-reddening study by Hodge & Lee (1988). The mean optical depth of six radial bins are given by the solid squares, with the error bars representing the statistical uncertainty ($= $ dispersion/$\sqrt{N-1}$).

**Figure 5.** Radial distribution of the optical depth to HI column density ratio $\tau_V/N(HI)$. Crosses are results for M31 cells. Solid squares with error bars are means for cells in six bins, each spanning 2 kpc in the plane of M31. The error bars give the one-$\sigma$ dispersions. The solid line represents the linear regression calculated for the sample of cells (358 of them). The dotted curve is a smooth fit to the means, given by Eq(18) in the text. The dashed horizontal line indicates the $\tau_V$–to–HI-gas ratio in Solar Neighborhood taken from Savage & Mathis (1979).

**Figure 6.** Comparison of the radial distribution of the $\tau_V$–to–HI-gas ratio obtained in this work with the results of Bajaja & Gergeley (1977) and those of Walterbos & Kennicutt (1988, WK88).

**Figure 7.** Plot of the 'gradient-corrected' optical depth versus HI column density. Cells in different part of the disk are denoted by different symbols. The solid line represents the linear regression of the data.

**Figure 8.** Radial distribution of the contribution to the FIR emission due to the heating of diffuse dust by the UV (912 — 3000Å) radiation and by the optical (3000 — 9000Å) radiation. Plotted are also the sum of the two contributions ('total heating'), and the observed FIR surface brightness distribution.

**Figure 9.** Comparison of radial distributions of $\tau_V$–to–$N_{HI}$ ratio of cells in the northwest half (near-half) and in the southeast half (far-half). The error bars of the means represent the statistical uncertainties of the means.

**Figure 10.** Comparison of the dust–to–HI-gas mass ratio estimated using our model ($M_{d,V}/M(HI)$) and that estimated from the FIR emission ($M_{d,100}/M(HI)$). Both are normalized by the value in Solar Neighborhood. Cells in different part of the M31 disk are denoted by different symbols.